\documentclass[aps,prl,superscriptaddress]{revtex4}

\usepackage{amsmath}
\usepackage{graphicx}
\usepackage{MnSymbol}

\begin{document}

\title{Comment on ``On the Origin of Frictional Energy Dissipation''}

\author{B.N.J. Persson}
\affiliation{PGI-1, FZ J\"ulich, Germany}
\affiliation{MultiscaleConsulting, Wolfshovener str 2, 52428 J\"ulich, Germany}

\begin{abstract}
In their interesting study (Ref. \cite{Hu}) Hu et al have shown that 
for a simple ``harmonium'' solid model the slip-induced
motion of surface atoms is close to critically damped. This result is in fact well known from 
studies of vibrational damping of atoms and molecules at surfaces. However,  
for real practical cases the situation may be much more complex and the conclusions of Hu et al invalid.  
\end{abstract}

\maketitle

\begin{figure}
\centering
\includegraphics[width=0.7\textwidth]{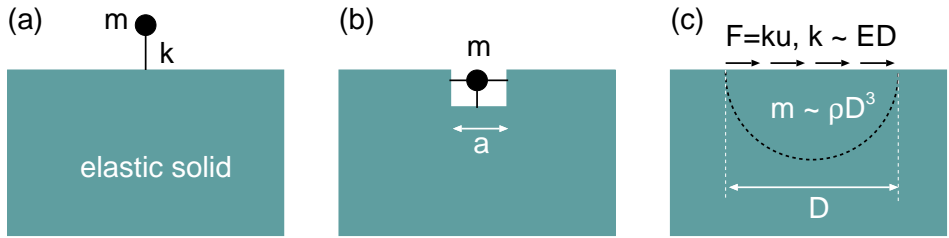}
\caption{\label{3block.eps}  
(a) An atom (mass m) bound to an elastic half space, with the bond spring constant $k$.
(b) An atom in the surface plane of the solid bound to the surrounding atoms with bonds
with the spring constant $k$. (c) A stress $\tau$ acting on the surface of an elastic half space,
within a circular region of diameter $D$, gives rise to a lateral displacement $u$ given by
$F=ku$, where $F$ is the total lateral force and\cite{spring} $k\approx ED$ ($E$ is the Young's elastic modulus)the contact stiffness.}
\end{figure}

Consider first the simplest case of an atom (mass $m$) adsorbed on an elastic half space
[see Fig. \ref{3block.eps}(a)]. Assume that the bond
can be described by an (harmonic) elastic spring with the spring constant $k$. 
If the substrate would be rigid the equation of motion
for the displacement coordinate (parallel or normal to the surface) $u$ is
$$m\ddot u = - k u,\eqno(1)$$
so the atom would vibrate (undamped) with the frequency $\omega_0 = (k/m)^{1/2}$.
However, if the substrate has a finite elasticity (Young's modulus $E$) 
the adsorbate motion will be damped due to emission of 
elastic sound waves (phonons). Using the elastic continuum model one can show that\cite{P1,P2}
$$m\ddot u = - k' u - m\eta \dot u\eqno(2), $$
where the damping $\eta$, due to emission of phonons into the elastic half space, is
$$\eta \approx {m\over \rho} \left ( {\omega_0 \over c}\right )^3 \omega_0,\eqno(3)$$
where $\rho$ is the solid mass density and $c \approx (E/\rho)^{1/2}$ the sound velocity.
Thus the interaction with the substrate will result in a damping $\eta$ of the vibrational motion, and also
to a renormalization of the spring constant $k\rightarrow k'$, but we will neglect the latter effect.  If we assume
$u\sim {\rm exp}(-i\omega t)$ then from (2)
$$\omega = -i {\eta \over 2} \pm \left (-{\eta^2 \over 4}+\omega_0^2\right )^{1/2}$$
The motion will be critically damped if $\eta \approx 2 \omega_0$.

We can apply (3) to the motion of a block atom as in Fig. \ref{3block.eps}(b) or to a surface patch as in Fig. \ref{3block.eps}(c). 
In this case the spring constant\cite{spring} $k\approx E D$, where $D$ is the diameter of the patch (for the atom case $D=a$ is the
lattice constant) and the effective mass $m \approx \rho D^3$. Substituting this in (3) and using that the sound velocity
$c \approx (E/\rho )^{1/2}$ we get 
$$\eta \approx {c\over D}$$ 
and 
$$\omega_0 \approx \left ( {k\over m} \right )^{1/2} \approx \left ({E D \over \rho D^3}\right )^{1/2} \approx {c \over D}$$
Hence $\eta \approx \omega_0$ and the motion of the surface patch (or when $D=a$ the atom) is nearly critical damped, as also found
by Hu et al in their detailed lattice dynamics calculations. Note that $1/\eta \approx D/c$ is just the time it takes for the elastic
wave to propagate the distance $D$ needed to remove the vibrational energy from the volume element $D^3$.

In vibrational spectroscopy, the process described above is referred too as vibrational energy relaxation
because the energy put initially into the (quasi) localized vibrational mode is transferred to the surrounding. 
However, Hu et al consider the full system and instead refer to the process as phase relaxation since the
initial state can be considered as originating from a very special superposition of crystal vibrational eigenmodes
with definite phases; with increasing time the phases changes and the amplitude of the vibrational excitation is
no longer localized to its original spatial location. In adsorbate vibrational spectroscopy, phase relaxation
instead refers to the influence of the fluctuating (due to the irregular thermal motion) environment 
on the phase of the (quasi) localized vibrational mode. 

The conclusion that the slip motion is nearly critically damped may not hold for real (practical) systems for the following reason: 
Real solids will almost always have modified surface layers (e.g. oxides on metals) and contamination films. In this case long-lived 
(strongly under-damped) vibrational excitations may occur localized to the contamination film or modified surface layer, which could
strongly affect the frictional properties of the solid. Long lived vibrtational excitations are very well known from vibrational
spectroscopy of adsorbed molecules. Thus, if the vibrational frequency $\omega_0$ is above the top of the bulk phonon band the vibrational
mode can decay (via phonon excitation) only via multi-phonon processes determined by the anharmonic adsorbate-substrate interaction.
This can result in extremely long vibrational lifetimes. One extreme (and well studied) case is carbon monoxide adsorbed on a NaCl
crystal\cite{Ew}: The C-O stretch vibration frequency is $\approx 8$ 
times higher than the highest NaCl phonon frequency, so a multiphonon
emission process involving at least 8 bulk phonons would be necessary for decay by phonon emission. The probability for such a
process is extremely small and in fact the exited C-O molecule on NaCl is know to decay mainly by photon emission (infrared fluorescence) 
resulting in a very long vibrational relaxation time $\tau \approx  4 \ {\rm ms}$. 
Another well studied case involves silicone surfaces passivated by
adsorption of hydrogen atoms. The Si-H vibration frequency is well above the top of the bulk phonon band resulting again in a very long lived 
vibrational excitation.

Most real surfaces have also layers of weakly adsorbed molecules, e.g., hydrocarbons. In these cases very low-frequency vibrations may also occur
and these too are weakly damped. This follows from (3): weakly bound molecules result in very low frequency modes (frustrated translations) 
and since $\eta \sim \omega_0^4$ this will result in very long vibrational lifetimes. 
Thus, for example, for saturated hydrocarbones on 
metallic surfaces, or on hydrogen terminated diamond, for the perpendicular molecule-substrate 
vibrations typically\cite{P3} $\eta/\omega_0 \approx 0.05$, and for the
parallel vibrations (which may be more relevant for friction) $\eta/\omega_0$ 
is even much smaller (vibrational lifetimes of order ns); in these cases the damping $\eta$ may, on metals, 
be dominated by excitation of electron-hole pairs rather than by phonon emission\cite{P3}.

For a nanosized asperity contact regions (contact diameter $d$), 
the contact time may be typically $d/v$, and for macroscopic friction experiments where
$v$ may be of order $1 \ {\rm m/s}$ this gives contact times of order 1 ns. It is clear that in such experiments (non-thermal) 
vibrational excitations may occur at the sliding interface which could influence the friction dynamics.


\begin{thebibliography}{99}
\bibitem{P1}
A rigorus derivation of (2) and (3) was given in Appendix B in Ref. \cite{P2}.
A simple ``back of the envelope'' derivation of (3) is given on page 183 in \cite{P3}.
\end{thebibliography}
\end{document}